\documentclass{aa}
\usepackage{graphicx}
\usepackage{rotating}

\begin{document}
\title{ 3D spectroscopy with VLT/GIRAFFE: I- the true Tully Fisher relationship at z$\sim$ 0.6  \thanks{based on FLAMES/GIRAFFE Paris Observatory Guaranteed Time Observations collected at the European Southern Observatory, Paranal, Chile, ESO Nos 71.A-0322(A) and 72.A-0169(A).} }

\author{  H. Flores\inst{1}
          \and 
          F. Hammer\inst{1}
          \and
          M. Puech\inst{1}
          \and 
          P. Amram\inst{2}
          \and
          C. Balkowski\inst{1}
          }

\offprints{hector.flores@obspm.fr}
\authorrunning{H. Flores et al.}
\titlerunning{Tully Fisher relationship at z$\sim$ 0.6}

   \institute{Laboratoire Galaxies Etoiles Physique et
        Instrumentation, Observatoire de Paris-Meudon, 5 place Jules Janssen,
        92195 Meudon France 
         \and
Observatoire Astronomique
Marseille-Provence \& Laboratoire d'Astrophysique de Marseille, 2 Place Le Verrier, 13248 Marseille Cedex 04, France.
       }
\date{---}

\abstract{\\
 A precise derivation of the evolution of the Tully Fisher is crucial
  to understand the interplay between dark matter and baryonic
matter in cosmological models, 
 Using 15 deployable integral field
 units of FLAMES/GIRAFFE at VLT, we have recovered the velocity fields
 of 35 galaxies at intermediate redshift (0.4 $<$ z $<$ 0.75). This
 facility is able to recover the velocity fields of almost all the
 emission line galaxies with I$_{AB}\le$22.5 and
 $W_0$(OII)$\ge$15\AA.
 In our sample, we find only 35\%  rotating
 disks. These rotating disks produce a Tully-Fisher relationship
 (stellar mass or $M_K$ versus V$_{max}$) which has apparently not
 evolved in slope, zero point and scatter since z=0.6. The only
 evolution found is a brightening of the B band luminosity of a third
 of the disks, possibly due to an enhancement of the star
 formation.    
 The very large scatters found in previously reported
 Tully-Fisher relationships at moderate redshifts are caused by the
 numerous (65\%) galaxies with perturbed or complex kinematics.
 Those galaxies include minor or major mergers, merger remnants and/or
 inflow/outflows and their kinematics can be easily misidentified by
 slit spectroscopy. Their presence suggests a strong evolution in the
 dynamical properties of galaxies during the last 7 Gyrs.
 }

\maketitle
%

\section{Introduction} 
The relationship between the luminosity and HI maximal rotation
velocity discovered by Tully and Fisher (1977, hereafter called TF)
was first used to estimate distances of late-type galaxies. Widely
studied in the local Universe at different optical and near infrared
bands, this relationship always showed a tight and linear
correlation. Corrections to fundamental observables have been
established by Tully and Fouqu\'e (1985) and revisited by Tully et
al. (1998), especially concerning the global extinction effects.
Verheijen (2001) compared B, R, I and K band luminosities and HI
rotation curves and concluded that the TF relation reflects a fundamental
correlation between the light and the mass of the dark matter halo.
Karachentsev et al. (2002) using 2MASS survey data has confirmed this
result and found a TF slope increases steadily from the B to K band.

Since 1997, many attempts have been made to derive the TF relationship
at higher redshifts and to investigate its possible evolution. Vogt et
al. (1997) and Simard \& Pritcher (1998) obtained the first
rotation curves of disk-like galaxies using long slit observations aligned with the
optical axis, which have been modelled using deconvolution
methods. The results of these studies are quite controversial: while
Vogt et al. (1997) found only a minor evolution of the TF relation
(an average effect of less than 0.2 mag in absolute B-band luminosity),
Rix et al. (1997) and Simard \& Pritchet (1998) found a strong
brightening of roughly 2 mag. Later, Ziegler et al.  (2002) and Bohm
et al. (2004) using the FORS Deep Field (FDF) survey also found a
significant evolution in the B band: at intermediate redshifts, they claim
that the TF relationship shows a flatter slope than that of local
galaxies.
 Ferreras \& Silk (2001) and Ferreras et al. (2004) have modelled the
change of the TF slope in the B band, which has been interpreted as an
enhancement of the star formation. More recently, Conselice et
al. (2005) have combined Keck spectroscopy and near-infrared
imaging, and have investigated the K-band and stellar mass TF
relation. They found a lack of evolution in either the K-band or
stellar mass TF relation.

However, all the above studies at intermediate redshifts
reveal extremely scattered Tully-Fisher relationships,
i.e. they show scatters that are several magnitudes larger than that of the tight
relationships for local galaxies.  It is possible that this scatter at
high z is the explanation of most of the discrepancy between the
various studies, and it is crucial to understand whether the
 TF relationship is preserved at higher redshift, or if it is only
a property of local galaxies.

We notice that all the former investigations of the Tully-Fisher
relationship at intermediate redshift were based on long (or multi)
slit spectroscopy, and thus the large dispersion may be related
to instrumental effects.  Rotation curve anomalies such as those
detected by Kannappan et al. (2002) using long-slit spectroscopy of
nearby galaxies can be explained by velocity field disturbations due
to minor or major mergers during the history of each galaxy, but it is
unclear whether such details can be identified by using long slit
spectroscopy in distant galaxies. It is now urgent to investigate the
full 3D kinematics of emission line galaxies, and to identify possible
disturbances on rotation curves, before producing thousands of
rotation curves with slit spectroscopy. FLAMES/GIRAFFE offers a unique
possibility to ``revisit'' the TF relationship at moderate redshift
with a more powerful tool, i.e. a multi-integral field unit
spectrograph with a spectral resolution of about 10000\AA.


This is the first paper of a series of three articles dealing with the
results of observations in the frame of FLAMES/GIRAFFE Paris
Observatory Guaranteed Time. This paper describes the observations,
sample selection, methodology and first results. Paper II
(Puech et al. 2006a) discusses in detail the kinematics of the most
compact galaxies of the sample. Paper III (Puech et al. 2006b)
presents the electron density maps for six galaxies of our sample.

This paper is organized as follows: Section 2 describes the
observations and the selection of the sample ; Section 3 shows the
methodology used to reconstruct velocity fields and $\sigma$-maps and a
simple scheme to classify them; in Section 4 we provide the B and K
absolute band TF relations, followed by a detailed discussion in
Section 5. We adopt the $\Lambda$CDM cosmological model ($H_0$=70 km
s$^{-1}$ Mpc$^{-1}$, $\Omega _M$=0.3 and $\Omega _\Lambda =0.7$) in
this paper.

\section{Observations}
A total of 5 nights were allocated during the FLAMES/GIRAFFE
guaranteed time to study the kinematics of distant galaxies (ESO runs
071.B-0322(A) and 072.A-0169(A)). We used the IFU mode (3''x2'' array
of 20 squares 0.52 arcsec/pixel microlenses) with setups LR04
(R=0.55\AA -- 30 km/s) and LR05 (R=0.45\AA -- 22 km/s), and
an integration time ranging from 8 to 13 hours (see Table
1). Observational seeing during individual exposures  ranged from
0.35 to 0.8 arcsec. To prepare OzPoz configuration files including
guide star and fiducial fibers (used to center the plate), relative
astrometry for each observed field was prepared using reduced 12k
images (CFHT public database: CFRS03hr and CFRS22hr) and the HDFS
public Goddard images combined with UCAC and USNO-B catalogs.  The
final accuracy of the astrometrical solution is better than 0.2 arcsec
in rms for most galaxies. The same astrometrical solution
was used for each exposure (OzPoZ pointing error$<$ 10$\mu$m).

Data reduction was done using the dedicated BLDRS software
developed at the Geneva Observatory (Bl\'echa et al., 2000). In order
to verify the fiber to fiber wavelength calibration, we systematically
control the wavelength of two sky lines in each fiber of the IFU. We
found a relative error per IFU bundle $\le$ 0.2 km/s. For reduced
spectra, two methods were used to subtract the sky, using the standard
software option or manually using an IDL dedicated task.  Data cubes
using the [OII] doublet emission lines for each galaxy were
constructed using our IDL dedicated package.

{\scriptsize
\begin{table}
\centering
\begin{tabular}{lccr}\hline
Run ID & Field & Setup & Exp time (hr) \\\hline
071.B-0322(A) &  CFRS03hr & L05 &13hr\\
071.B-0322(A) &  HDFS     & L04 & 8hr\\
071.B-0322(A) &  HDFS     & L05 & 8hr\\
072.A-0169(A) &  CFRS03hr & L04 & 8hr\\
072.A-0169(A) &  CFRS22hr & L04 & 8hr\\\hline
\end{tabular}
\caption{Table of observations}
\label{tab1}
\end{table}
}

\subsection{Sample selection}
 Galaxies were selected from the CFRS survey and from the HDFS. 
Given the difficulty in selecting fifteen galaxies with [OII]3727 emission line in the small redshift range available with FLAMES/GIRAFFE (for example z=0.55-0.73 for the LR05 mode), some galaxies with faint emission were also included during the observation to test the capabilities of the FLAMES/GIRAFFE facility. Among the 60 observed galaxies,  45 galaxies have a redshift between 0.4 and 0.75 with EW$_0$([OII])$>$ 0. Among them,  three  (CFRS030560, CFRS030579 \& CFRS030728) were rejected due to misidentified emission in CFRS spectra, and one (CFRS031319) was rejected because of  strong contamination from a sky line. Six galaxies  (CFRS030032, CFRS030098, CFRS030717, CFRS030767, HDFS4010 \& HDFS4080) have  EW$_0$([OII])$<$ 15\AA$\;$  and we found that they do not have enough emission to be detected by FLAMES/GIRAFFE. Figure 1 shows the performances of FLAMES/GIRAFFE which can detect all galaxies having  EW$_0$([OII])$\ge$ 15\AA$\;$ after 8 to 13hr of integration time.  
Our final sample is then composed of  35 intermediate redshift galaxies having EW$_0$([OII])$\ge$ 15\AA$\;$ and I$_{AB}\le$22.5 (Table 2). Our selection process is representative of emission line galaxies of the CFRS, from z=0.4 to z=0.75.

\begin{figure}[h!]
\centering 
\includegraphics[width=8cm]{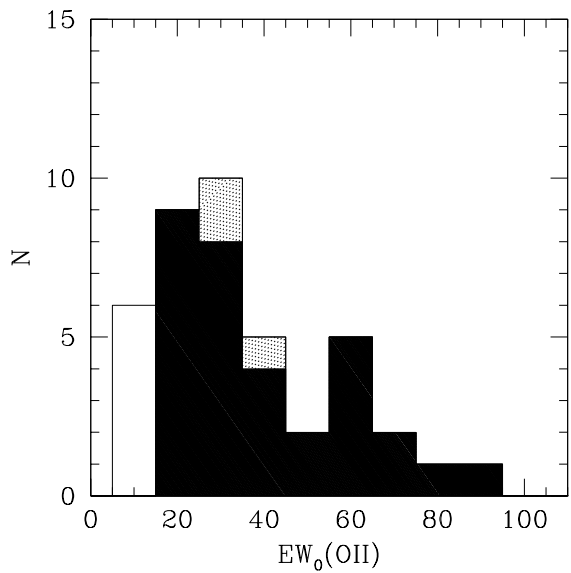}
\caption{EW$_0$([OII]) histogram for our sample galaxies. For galaxies with  EW$_0$([OII])$\ge$ 15\AA$\;$, GIRAFFE can retrieve their kinematics after 8hr of integration time (black region). The major difficulty  studying the detected galaxies is then related to the flux strength. The three galaxies with EW$_0$([OII])$\ge$ 15\AA, but with  too spatially concentrated emission are displayed in grey. The open histogram includes objects with EW$_0$([OII])$<$ 15\AA$\;$.}
\label{ew}
\end{figure}

\subsection{Detection of the [OII] doublet and the S/N}
The accuracy with which the velocity and the velocity dispersion can
be recovered from emission lines depends on by how the lines are higher
than the noise in the continuum. 
In kinematical studies of local galaxies, only spatial pixels with
$S/N > 1.5$ are considered (following Amram 1991), with the S/N defined as the
ratio between the flux barycenter and the pseudo-continuum
noise. However, this criterion cannot be directly used here because
the [OII] emission line is a doublet: the local minimum in the flux
between the two emission lines leads to very low values of the signal
to noise ratio. 
We choose to define the S/N as the ratio between the total flux in the
doublet and the noise in the pseudo-continuum (see equation below): this allows us to
exploit the detection of the two emission lines in the [OII] doublet.
We then normalised this S/N by the number of spectral elements of
resolution in the doublet, i.e. roughly half the total number of
spectral pixels over which the [OII] emission line is spread. This
gives the mean S/N in the emission line:

$$
<S/N>= \frac{\sum^{N}_{i} S_i }{\sqrt(N) \times \sigma}
$$

\noindent
where N is the number of spectral resolution elements in which the [OII]
doublet is detected, $S_i$ is the signal in the pixel $i$ and $\sigma$
is the dispersion of the noise in the continuum of the spectra. In the
following, for simplicity, we will call this ``$<S/N>$'' simply
``spectral S/N'' or even ``S/N''.

After visual inspection of spectra at differents $S/N$, we decided to
define a mimimum quality criterion. We keep only datacubes containing
at least four GIRAFFE spatial pixels with S/N$\ge 4$. Among the 35
datacubes, three (CFRS030186, CFRS030327 \& CFRS030589) were rejected
by this criteria. These three galaxies have their emission too
concentrated to be observed with the spatial resolution of the GIRAFFE
IFU. Within the 32 remaining datacubes, only spatial pixels with
S/N$\ge$ 3 were kept to established the velocity field and the
$\sigma$-map (equivalent to one emission line with a $S/N > 6$). 
Among the 32 galaxies we found that
between 6 and 18 spatial pixels (with a median of 10 pixels) passed
this last selection.

\subsection{Morphological parameters}
\begin{figure}[h!]
\centering
\includegraphics[width=8cm]{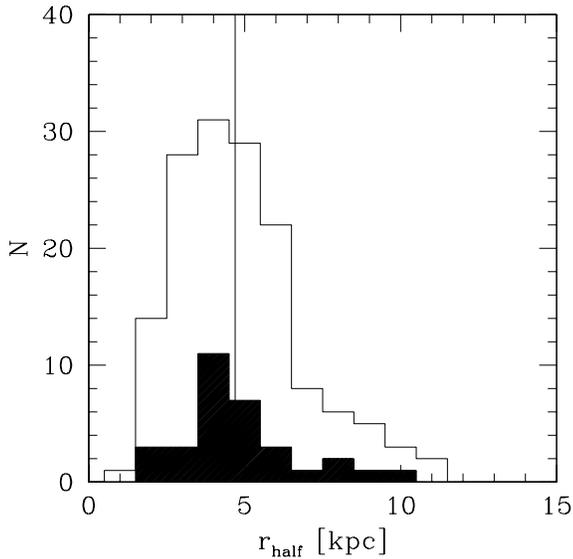}
\caption{Observed size distributions of the sample (black histogram) compared to the size of the CFRS/LDSS survey (open histogram). The vertical line shows the median size from the CFRS.
}
\label{half}
\end{figure}

Galaxies included in this sample display all the previously reported morphologies of intermediate redshift galaxies with emission lines, including spiral, irregular, compact and merger galaxies. For each galaxy we have estimated the half light radius (r$_{half}$) using two techniques (polyphot and ellipse software under IRAF, see Hammer et al. 2001 for a complete description of the procedure). 
The sample presents a median  r$_{half}$ radius of 4.7 kpc, which is close to the median size of CFRS galaxies (Figure \ref{half}, see also Lilly et al. 1998).  We found that $\sim$25\% of our sample have  r$_{half}$ $\le$ 3.5 kpc (Guzman et al. compactness criterion, see Figure \ref{half}) and 50\% satisfy the  LCG criterium (i.e., r$_{half}$ $<$ 4.75 kpc) from Hammer et al. (2001). Detailed properties of LCG kinematics are described in paper II (Puech et al, 2006a). The 35 galaxies selected here have morphologies and compactness comparable to that of field galaxies with emission lines  (see Zheng et al. 2004, 2005 and Hammer et al., 2005). 

Inclinations for each galaxy were derived from HST images (WFPC2, 0.1 arcsec/pixel or ACS, 0.05 arcsec/pixel, ESO/HST Public archive) or CFHT images for a few remaining cases (0.207 arcsec/pixel). By comparing results from Sextractor (Bertin \& Arnouts, 1996) to those from the ellipse task of IRAF, we estimated the mean error to be $\sim 2$ degrees. Independent measures were also made by eye and gave similar results with an estimated mean error of $\sim 4$ degrees when compared to the above methods. In the following, we will use the results from automatic methods and assume an error of $\pm 4$ degrees. 

\section{Kinematics of distant galaxies}
\subsection{Methodology}
Among the complete sample of 35 emission line galaxies observed with
GIRAFFE, 32 galaxies are either presented in Figure 3 (15 with
r$_{half}$ $>$ 4.75 kpc) or in Figure 1 of paper II (17 compact galaxies, Puech et al., 2006a). In both Figures we have used a S/N mask described
in section 2.1 to show their velocity fields and $\sigma$ maps. These
galaxies show on average 10 spatial GIRAFFE pixels for
which the $S/N > 0.75$ for
the [OII] emission lines.
\setcounter{figure}{3}
\begin{figure*}[ht!]
\centering
\includegraphics[width=16cm]{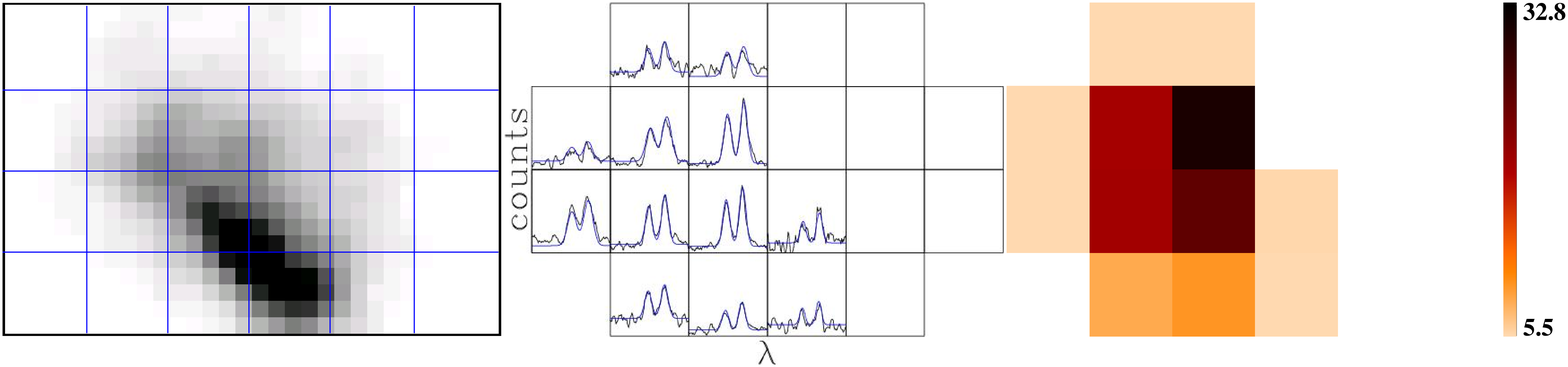}
\caption{ Example of the fitting procedure on the galaxy CFRS030488 at z=0.601, for which the [OII]$\lambda$3729,3726\AA$\;$ doublet is well resolved. The example shows the microlens grid superimposed to the galaxy (left panel), each microlens (spatial pixel) providing a spectrum (central panel). Twelve spectra were selected using our S/N criteria (right panel) and are shown superimposed with the best fit (blue line).  
}
\label{resol}
\end{figure*}

The [OII] doublets were selected by visual inspection using a
Savitzky-Golay filtering which has the advantage over the widely used
box smoothing average method of conserving the first momentum of
spectral lines (Press et al. 1989). We then fitted a double Gaussian
with the following constraints: $\lambda _2$-$\lambda _1$= 2.783$\AA$
at rest and $\sigma_1$=$\sigma _2$. The line ratio was free except
when the fit failed: we then forced the line ratio to 1.4 which was
the median value observed in our integrated spectra. This is caused
either by the low S/N, or by the difficulty to deblend the [OII]
doublet in some complex systems (see Figure 4 and below). It affects
only $\sim$ 10-15 \% of the measured pixels and thus cannot
significantly alter our conclusions. In each case, we have checked by
eye if the derived fit was acceptable. In some cases, we found blended
[OII] spectra, in spite of the relatively high spectral resolution
($R\sim 10000$). This effect is independant of the S/N and is thus not
an instrumental/observational artefact (see also paper II, Puech et
al. 2006a). This could be due to high local velocity gradients,
multiple structures in the velocity field and/or high local
extinction. Most complex kinematics show blended doublets rather than
 more regular (rotating disks and perturbed) kinematics (see next
section). 

For each galaxy, we have computed a set of 3 maps (see Figure
\ref{sample1}): velocity fields, velocity dispersion maps
($\sigma$-maps in the following) and S/N maps. To make  the
interpretation easier we present velocity fields and $\sigma$-maps
using a simple 5x5 linear interpolation (see figure 5). Velocity
fields have been derived after subtraction of the $\sigma$-clipped
mean. Sigma maps have been corrected for instrumental dispersion
measured on a sky line. Because GIRAFFE is a very stable instrument,
we have  neglected the variation of the instrumental response with
wavelength.  The velocity and velocity dispersion maps were
constructed independently by two members of our team (HF and MP), and
then compared. We estimated the error in the derived velocities and
velocity widths by taking into account both the difference between
these two analyses and the error due to the S/N. For the velocity, we
found an error $\Delta V < 5 km/s$ for spectra with $4 >S/N \geq 3$. 
For the velocity width, we found an error
$\Delta \sigma$ of 12 , 6 and $<5$ km/s for spectra with
$4 >S/N \geq 3$, $6 >S/N \geq 4$ and $S/N \geq 6$, respectively.

Table \ref{tab2} lists the properties of 32 galaxies used for our
analysis (column 1). The position is given (columns 2 \& 3), as well as the
redshift, the I band magnitude and the B and K absolute magnitude
(columns 4, 5, 6 \& 7), the stellar mass (column 8), the image origin (column
9), the morphological parameters r$_{half}$, inclination and
compactness classification (columns 10, 11 and 12), the V$_{max}$ and the kinematical classification (columns 13 \& 14).

\setcounter{figure}{2}
\begin{figure*}[h!]
\centering
\caption{SEE ATTACHED JPG FIGURE Optical images, velocity fields, $\sigma$ and S/N maps of large galaxies classified as rotating disks. Similar maps for more compact galaxies are presented in paper II (Puech et al., 2006a). Each spatial GIRAFFE pixel  has been simply 5x5 interpolated for visualisation (in the velocity fields and dispersion maps). For each galaxy the valid spatial pixels are indicated on the SN map (4th column). Optical images from CFHT are marked with an asterisk. }
\label{sample1}
\end{figure*}
\setcounter{figure}{2}
\begin{figure*}[h!]
\caption{SEE ATTACHED JPG FIGURE Cont. Optical images, velocity fields, $\sigma$ and S/N maps of large galaxies classified as perturbed
rotations or complex kinematics. They show either a rotation with a complex sigma map, or a complex sigma map and velocity field. Optical images from CFHT are marked with an asterisk.}
\label{sample2}
\end{figure*}
\setcounter{figure}{4}
\renewcommand{\baselinestretch}{1.0}
{\scriptsize
\begin{sidewaystable*}
\caption{Main properties of the sample of distant galaxies: galaxy
names, redshifts, isophotal I magnitudes, absolute B and K magnitudes,
origin of images, half light radii, inclinations and
compactness flags (1=compact).
}
\begin{tabular}{lccccccccccrrrr}\hline\hline
ID      & RA$^a$ & DEC$^a$    & z      & I$_{AB}^b$ &  M$_B$(AB)$^c$  &  M$_K$(AB)$^c$  & log(M$_{\star}$)$^d$ & image & r$_{half}^e$ & i$^f$ & Cm$^g$ & V$^h$ & Cl$^i$ \\\hline
CFRS030046& 03:02:28.716 & +00:13:33.94   & 0.5120 & 20.79      &  -19.75(-20.68) &  -21.67(-23.62) &10.50 &hst   & 8.30  & 66   & 0& 193.99 &  RD \\
CFRS030085& 03:02:25.285 & +00:13:25.13   & 0.6100 & 22.00      &  -19.63(-20.67) &  -20.82(-22.78) &10.08 &hst   & 7.59  & 71   & 0& 160.48 &  RD \\
CFRS030619& 03:02:46.942 & +00:10:32.63   & 0.4854 & 20.80      &  -20.66(-20.87) &  -21.93(-23.77) &10.50 &hst   & 3.87  & 27   & 1& 186.19 &  RD \\
CFRS031353& 03:02:48.407 & +00:09:16.54   & 0.6340 & 21.35      &  -20.45(-21.17) &  -22.42(-24.33) &10.72 &hst   & 6.40  & 57   & 0& 249.49 &  RD \\
CFRS039003& 03:02:32.159 & +00:06:39.14   & 0.6189 & 20.77      &  -21.24(-21.49) &   99.99( 99.99) &99.99 &hst   & 4.87  & 29   & 0& 277.34 &  RD \\
CFRS220504& 22:17:58.074 & +00:21:37.02   & 0.5379 & 21.02      &  -20.51(-20.86) &  -21.36(-23.22) &10.30 &cfh   & 4.51  & 42   & 0& 175.94 &  RD \\
CFRS221119& 22:17:41.448 & +00:18:54.45   & 0.5138 & 20.07      &  -21.25(-21.52) &   99.99( 99.99) &99.99 &cfh   & 4.79  & 31   & 0& 245.01 &  RD \\
HDFS4020  & 22:32:56.074 & -60:31:48.83   & 0.5138 & 21.13      &  -20.57(-20.97) &  -20.11(-21.98) & 9.82 &hst   & 4.99  & 50   & 0& 137.99 &  RD \\
HDFS4170  & 22:32:45.557 & -60:34:18.84   & 0.4602 & 20.79      &  -20.42(-20.95) &  -22.60(-24.48) &10.82 &hst   & 3.56  & 51   & 1& 207.37 &  RD \\
HDFS4180  & 22:32:58.008 & -60:35:25.95   & 0.4647 & 21.42      &  -20.13(-20.82) &  -20.38(-22.28) & 9.90 &hst   & 5.37  & 64   & 0& 137.83 &  RD \\
HDFS5190  & 22:33:00.088 & -60:35:29.91   & 0.6952 & 21.31      &  -21.25(-21.95) &  -21.92(-23.83) &10.51 &hst   & 4.01  & 59   & 1& 201.07 &  RD \\\hline
CFRS031032& 03:02:38.736 & +00:06:11.49   & 0.6180 & 20.49      &  -21.18(-21.46) &  -22.63(-24.48) &10.87 &acs   & 1.79  & 37   & 1& 166.50 &  PR \\
CFRS031349& 03:02:49.099 & +00:10:02.13   & 0.6155 & 20.87      &  -21.18(-21.73) &  -22.91(-24.80) &10.91 &hst   & 3.84  & 48   & 1& 282.52 &  PR\\
CFRS220321& 22:18:02.930 & +00:14:27.11   & 0.4230 & 20.93      &  -19.97(-20.34) &  -20.79(-22.65) &10.08 &cfh   & 5.36  & 42   & 0& 196.14 &  PR \\
CFRS220619& 22:17:54.573 & +00:18:59.52   & 0.4676 & 21.55      &  -19.33(-19.81) &  -19.32(-21.20) & 9.50 &hst   & 4.31  & 68   & 1&  80.45 &  PR \\
CFRS221064& 22:17:43.081 & +00:15:07.67   & 0.5383 & 22.08      &  -19.87(-20.27) &  -21.64(-23.50) &10.39 &hst   & 2.36  & 48   & 1& 149.13 &  PR \\
HDFS4040  & 22:32:52.742 & -60:32:07.26   & 0.4650 & 21.76      &  -19.89(-20.26) &  -20.04(-21.90) & 9.75 &hst   & 4.25  & 51   & 0& 116.85 &  PR \\
HDFS5150  & 22:33:02.454 & -60:33:46.48   & 0.6956 & 22.36      &  -20.19(-20.43) &  -21.02(-22.86) &10.14 &hst   & 3.38  & 42   & 1&  92.14 &  PR \\\hline
CFRS030488& 03:02:42.193 & +00:13:24.35   & 0.6069 & 21.58      &  -20.37(-20.55) &  -20.83(-22.67) &10.07 &hst   & 6.35  & 41   & 0&  66.48 &  CK \\
CFRS030508& 03:02:40.450 & +00:13:59.36   & 0.4642 & 21.92      &  -19.60(-19.82) &  -20.34(-22.18) & 9.87 &hst   & 3.32  & 38   & 1&  93.14 &  CK \\
CFRS030523& 03:02:39.375 & +00:13:27.14   & 0.6508 & 21.31      &  -20.66(-20.93) &  -21.54(-23.39) &10.35 &hst   & 3.57  & 41   & 1& 115.90 &  CK  \\
CFRS030645& 03:02:45.629 & +00:10:27.95   & 0.5275 & 21.36      &  -20.30(-20.66) &  -21.34(-23.20) &10.27 &hst   & 4.57  & 45   & 1& 149.85 &  CK   \\
CFRS031016& 03:02:41.006 & +00:06:55.45   & 0.7054 & 22.35      &  -19.58(-19.25) &  -21.23(-22.99) &10.23 &cfh   & 8.82  & 67   & 0&  19.55 &  CK \\
CFRS031309& 03:02:52.028 & +00:10:33.36   & 0.6170 & 20.62      &  -21.03(-21.77) &  -22.90(-24.81) &10.90 &hst   & 9.56  & 71   & 0& 106.44 &  CK   \\
CFRS220293& 22:18:03.560 & +00:21:31.27   & 0.5420 & 22.01      &  -19.69(-20.06) &  -20.89(-22.75) &10.10 &cfh   & 7.05  & 45   & 0& 150.03 &  CK \\
CFRS220919& 22:17:46.458 & +00:16:53.03   & 0.4738 & 21.77      &  -19.98(-20.18) &  -19.53(-21.36) & 9.54 &hst   & 2.52  & 60   & 1&  53.47 &  CK \\
CFRS220975& 22:17:45.117 & +00:14:46.71   & 0.4211 & 20.21      &  -20.40(-21.13) &  -22.52(-24.44) &10.82 &hst   & 3.82  & 50   & 1& 490.31 &  CK \\
HDFS4070  & 22:32:58.228 & -60:33:31.40   & 0.4230 & 22.04      &  -19.26(-19.43) &  -19.67(-21.50) & 9.62 &hst   & 3.23  & 40   & 0&  66.47 &  CK \\
HDFS4130  & 22:32:41.484 & -60:35:16.13   & 0.4054 & 20.09      &  -20.90(-21.19) &  -22.12(-23.98) &10.62 &hst   & 4.03  & 36   & 1& 184.09 &  CK \\
HDFS5030  & 22:32:57.517 & -60:33:05.94   & 0.5821 & 20.40      &  -21.73(-21.88) &  -22.68(-24.51) &10.81 &hst   & 4.19  & 25   & 1&  88.66 &  CK \\
HDFS5140  & 22:32:56.082 & -60:34:14.05   & 0.5649 & 22.38      &  -19.76(-20.34) &  -20.46(-22.35) & 9.91 &hst   & 2.56  & 50   & 1& 268.17 &  CK \\
HDFS4090  & 22:32:54.053 & -60:32:51.58   & 0.5162 & 22.15      &  -19.70(-19.69) &  -19.82(-21.63) & 9.67 &hst   & 1.53  & 45   & 1&  23.66 &  CK \\
\end{tabular}
\begin{list}{}{}
\item[$^{\mathrm{a}}$] J2000 coordinates.
\item[$^{\mathrm{b}}$] Isophotal magnitudes.
\item[$^{\mathrm{c}}$] Absolute AB magnitude following  the Hammer et al. (2005) method (uncorrected for dust). In parenthesis the corresponding Vega magnitudes corrected for dust/inclination (see section 4) are given.
\item[$^{\mathrm{d}}$] Stellar mass in M$_\odot$, following the Hammer et al. 2005 method .
\item[$^{\mathrm{e}}$] Half light radius in kpc.
\item[$^{\mathrm{f}}$] Inclination in deg.
\item[$^{\mathrm{g}}$] Morphological compactness following the Hammer et al. (2001) criteria (1=compact galaxy).
\item[$^{\mathrm{h}}$] Maximal gradient of the velocity field corrected from inclination and from instrumental effect (see section 3) in $[km/s]$.
\item[$^{\mathrm{i}}$] Kinematical classification (see section 3.2 for details): RD: rotating disks; PR: perturbed
rotations; CK: complex kinematics.
\end{list}
\label{tab2}
\end{sidewaystable*} 
}
\renewcommand{\baselinestretch}{1.0}

\subsection{Kinematical classification}

Using optical imagery and IFU observations we have defined 3 kinematical classes:
\begin{itemize}
\item {\bf Rotating disks (RD):} The axis of  rotation in the velocity field  follows the optical major axis and the $\sigma$-map shows a peak near the dynamical center ($\sigma$ maps should show a clear peak near the galaxy center where the gradient of the rotation curve is the steepest, e.g. Van Zee \& Bryant 1999);
\item {\bf Perturbed rotations (PR):} The axis of rotation in the velocity field follows the optical major axis, and the $\sigma$-map shows a peak shifted off the centre by at least 2 spatial GIRAFFE pixels, or does not show any peak;
\item {\bf Complex kinematics (CK):} Systems with  both velocity field and $\sigma$-map discrepant to normal rotation disks, including velocity fields  not aligned with the optical major axis (see  Figure 3 for more details). 
\end{itemize}

This simple classification scheme was chosen because the low
spatial resolution of GIRAFFE does not allow to identify small scale
structures as can be done from studies of nearby
galaxies. In the next section, we describe the status of each
individual galaxy presented in Figure 3.  Additional
comments and maps of galaxies classified as LCGs (Cm index in table 2) can be found in paper II  (Puech et al., 2006a).

\subsection{Notes on the 15 individual objects presented in Figure 3}
\noindent {\bf CFRS030046:} It is a rotating disk with a rotation
axis nearly parallel to its optical major axis. The $\sigma$ map
 peaks in the center. It has been classified as a Sbc galaxy by
Zheng et al. (2005).

\noindent {\bf CFRS030085:} This is an highly inclined galaxy  (71 degrees); its kinematics satisfies all the
criteria of a rotating disk. It is a luminous IR galaxy (LIRG, $L_{IR}$ $\ge$ $10^{11}$ $L_\odot$), which has been classified by Zheng et al.
(2004) as a spiral, although its color map resembles that of an
irregular galaxy.

\noindent {\bf CFRS031353:} It is almost a rotating disk with its rotation
axis parallel to its major axis and a peak in the center of its
$\sigma$-map. It has been classified as an Sab galaxy by Zheng et al.
(2005).

\noindent {\bf CFRS039003:}  It is almost a rotating disk with its rotation
axis parallel to its major axis and a peak in the center of its
$\sigma$-map. It has been classified by Zheng et al. (2004) as an Irr
galaxy with a large and red central region which is a possible bulge.
This galaxy is a LIRG with SFR $\sim$ 100 M$_\odot$/yr (from IR and
H$\alpha$ fluxes). Its electron density map has been recovered in paper III (Puech et al. 2006b).

\noindent {\bf CFRS220504:} It is a rotating disk with its rotation
axis parallel to its major axis. Its $\sigma$-map shows a double
peak in two IFU spatial GIRAFFE pixels near its center, They are probably an effect of the low
spatial resolution of GIRAFFE.  This double peak is reproduced by
our simple rotational disk model (see next section). Unfortunately, we only have a deconvolved CFHT image for this galaxy.

\noindent {\bf CFRS221119:} It is a rotating disk with the rotation axis
parallel to its major axis and its $\sigma$-map shows an elongated
peak corresponding to two IFU spatial pixels, which may be due to the low
spatial resolution of GIRAFFE and/or a border effect. Unfortunately, we
only have a deconvolved CFHT image for this galaxy.

\noindent {\bf HDFS4020:} A rotating disk with the rotation axis
parallel to its major axis and its $\sigma$-map shows an elongated
peak corresponding to two IFU spatial pixels (reproduced by our model), which
may be due to the low spatial resolution of GIRAFFE and/or to a border
effect at the top left spatial pixel. Its morphology reveals a spiral-like
structure revealed by the HST image.

\noindent {\bf HDFS4180:} A rotating disk with its rotation axis
nearly parallel to its major axis and a peak in the center of its
$\sigma$-map. Its morphology reveals a standard spiral structure on
the HST image.

\noindent {\bf CFRS220321:} It has a perturbed rotation without any peak
in the center of its $\sigma$-map. The velocity gradient seems to be
slightly rotated from the major axis. Unfortunately, we only have a
deconvolved CFHT image for this galaxy.

\noindent {\bf HDFS4040:} It has a perturbed rotation without a peak in the  center of the $\sigma$-map. Its morphology reveals both a large
spiral structure and a large diffuse component around the galaxy,
which could be a possible ``relic'' of a previous merger event.

\noindent {\bf CFRS030488:} its kinematics is classified complex. The $\sigma$-map shows a high S/N peak on the left and it shows a
perturbed region on the velocity field (on the top), which corresponds to a diffuse
component on the HST image. This region is oriented towards an
emission line galaxy at the same redshift (CFRS030485 at $z= 0.606$)
which is off by $<$ 46 kpc and which could be a fly-by companion. Its
morphology has been classified as irregular by Brinchman et
al. (1998).

\noindent {\bf CFRS031016:} Its kinematics is classified complex
because both the velocity field and the $\sigma$-map appear
perturbed. For this galaxy, we only have a deconvolved CFHT image.

\noindent {\bf CFRS031309:} Its kinematics is classified complex
because both the velocity field and the $\sigma$-map appear
perturbed. This system is actually a chain of galaxies undergoing an
obvious merger event and a high star formation rate is detected both in
radio and in infrared (Flores et al. 2004). The FLAMES/GIRAFFE IFU
covers only the central part of the system.

\noindent {\bf CFRS220293:}  Its kinematics is classified complex; its $\sigma$-map shows no peak in the centre and the rotation axis is not well aligned with the optical axis and our perfect disk simulation can not reproduce the $\sigma$-map. Unfortunately, we have only  a deconvolved CFHT image for this galaxy.

\noindent {\bf HDFS4070:} Its kinematics is classified complex
because both its velocity field and $\sigma$-map appear perturbed. Its
morphology reveals a bar and a diffuse component which
extends beyond the size of the IFU.

{\scriptsize
\begin{table*}
\centering
\begin{tabular}{c||r|r|r}\hline
Number (\%)            & All        & LCGs        & non-LCGs \\\hline
Rotating disks         & 11 (34\%)  &  3 (18\%)   & 8 (53\%) \\
Perturbed rotation     &  7 (22\%)  &  5 (29\%)   & 2 (13\%) \\
Complex  kinematics    & 14 (44\%)  &  9 (53\%)   & 5  (33\%)\\
\end{tabular}
\caption{Kinematical classification of the sample of 32 galaxies based on the visual examination of velocity field and sigma maps, and the test of the classification (see section 3.4).}
\label{tab3}
\end{table*}
}

Combining optical imagery and 3D information from paper II (Puech et al, 2006a) and this paper, we find that only 34\% of galaxies
are rotating disks. Other galaxies show some kind of disturbance going from perturbed $\sigma$-map to complex velocity field and $\sigma$-map. Table 3 summarises the classification of our sample (see also section 3.4).

\subsection{Testing the classification}
We are aware that classification of galaxy kinematics is not an easy
process.We stress that the above classification has been made
independently by 3 of us (HF, FH and MP) and then compared. This is a similar problem to that of the classification of
morphologies of distant galaxies during the 80s, after the release of
the deep images of field galaxies by the HST.

To help in the morphological classification of galaxies, generic
software (such as GIM2D or GALFIT) have been developed.  Most of them
are based on the comparison to a standard model, e.g. a disk plus
bulge model. Any deviation from these standards can be taken as a
measure of the level of irregularity. We propose to adopt a similar
scheme, and then relate our observed velocity fields to the sigma
maps, assuming that both are related to a rotating disk.

In the following we then assume that ALL the observed galaxies are
indeed rotating disks and that all the observed large scale motions in
the velocity fields correspond to rotations. We can derive a pseudo
rotational curve for each galaxy. Accounting for instrumental PSF and
seeing effects, we have derived the corresponding sigma maps. In doing
so, we have tried to optimise the coincidence between the sigma peak
in the observed map to that of the simulated map, using the full range
of uncertainties in the estimations of the inclination and PA. In
other words, we have tried to force each system to appear as a
rotational disk.  Under this condition, we have re-scaled all the
sigma intensities of the peak in the simulated sigma map to those of the
observed map, at the same spatial GIRAFFE pixel location. Because the
amplitude of the sigma peak is directly proportional to the amplitude
of the maximal velocity, the re-scaling of sigma (model) to sigma
(observed) has been limited within the range of 1.0-1.5. The latter
value corresponds to an upper limit of the uncertainty on the estimate
of $V_{max}$. To simulate the sigma maps derived from velocity fields,
we have developed the following procedure for each galaxy:
\begin{itemize}
\item A high resolution velocity field covering the whole IFU FoV is
generated using a standard rotation curve parameterization with the following
assumptions: The inclination is determined from HST images,
the PA is identified with the direction of the
maximal gradient within the velocity field, the dynamical center is
 located at equal distance to the maximal and 
minimal velocities of the velocity field, and the amplitude of the
rotation curve is taken to be the measured 2V$_{max}$ in the GIRAFFE
velocity field, corrected for inclination (see section 3.5);

\item A high resolution datacube is then simulated assuming a simple
Gaussian shape for each emission line and using the following recipe:
the flux sum of the lines falling in a given spatial GIRAFFE pixel matches
the [OII] flux detected in the corresponding  pixel of the GIRAFFE
datacube (counts were distributed uniformly in each spatial GIRAFFE pixel), and
each line is supposed to have an intrinsic dispersion equal to the
minimal velocity dispersion of the GIRAFFE $\sigma$-map;

\item Using this high resolution datacube, a GIRAFFE datacube is
simulated, taking into account a 0.81 arcsec seeing (median value at
0.5 nm at Paranal). A $\sigma$-map is computed from this
simulated GIRAFFE datacube;

\item This procedure is iterated two times: during the first
iteration, the fluxes measured on the simulated GIRAFFE spectra are
reinjected as new GIRAFFE fluxes, and the process is repeated. During
the second iteration, a correction factor is applied to the maximal
velocity gradient to make the amplitudes of the simulated and observed
$\sigma$-maps match. This factor is required to account for the
influence of the seeing and the large pixel size of the GIRAFFE IFU;

\item A last iteration is done for some galaxies, trying to optimize
the PA within the uncertainty in the direction of the maximal velocity
gradient in the velocity field (both the maximal and minimal velocity
positions are uncertain at $\pm$ half a GIRAFFE pixel).
\end{itemize}

We use two parameters to compare simulated and observed $\sigma$-maps.
The first one is the distance (in spatial GIRAFFE pixels) separating
the location of the peak in both maps, $d(MAX_{obs},MAX_{mod})$. The
second one is the (normalised) difference between the amplitude of
the peak of the observed $\sigma$-map and of the amplitude of the
simulated $\sigma$-map measured at the same pixel location:
$\epsilon_\sigma[max obs]$ $=$ $(\sigma_{obs}[Max
obs]-\sigma_{mod}[Max obs])/\sigma_{mod}[Max obs]$.  Figure
\ref{indices} shows the distribution of galaxies in the
$\epsilon_\sigma[max obs]$ vs. $d(MAX_{obs},MAX_{mod})$ plane. In this
plot, blue dots, green squares and red triangles represent
respectively the galaxies classified as rotation disk, perturbed
rotation and complex kinematics (see Table 2). All rotation disks but
one (CFRS031349) have $\epsilon_\sigma[max obs]< 0.2$ and
$d(MAX_{obs},MAX_{mod})$ = 0. Conversely, we find one galaxy
(CFRS031032, for which the kinematics was formerly classified as complex),
which could be a perturbed disk, although the simulated sigma peak is
much larger than the observed one. Note also that this galaxy is the
most compact of the sample (see details in paper II, Puech et
al. 2006a).

Two galaxies (CFRS030508 and HDFS5140), for which
velocity fields  probably result from an outflow (the
kinematical axis is almost perpendicular to the major axis of the
galaxy), appear as rotational disks in Figure \ref{indices}. This
shows the limit of the above method since outflows mimic rotational
motion when considering the relation between velocity fields and sigma
maps. Those two objects are kept as having complex kinematics.

\begin{figure*}[h!]
\centering
\caption{SEE ATTACHED JPG FIGURE Comparison between observed and modeled $\sigma$-maps of four
galaxies. The two first are classified as rotating disks (CFRS039003
and CFRS220504) and the two next are of the PR and CK categories
(respectively, CFRS220293 and HDFS4070). The first two columns are the
observed raw maps (VF and $\sigma$), the third column is the observed
5x5 interpolated $\sigma$-maps and the last column shows the 5x5
interpolated $\sigma$-map obtained from the perfect rotating disk
model described in the text. The $\sigma$-map can be easily reproduced
for rotating disks, while the observed $\sigma$-map not can be traced
by our simple model for galaxies with more complex kinematics.}
\label{fakesigma}
\end{figure*}

\begin{figure}[h!]
\centering
\includegraphics[width=8cm]{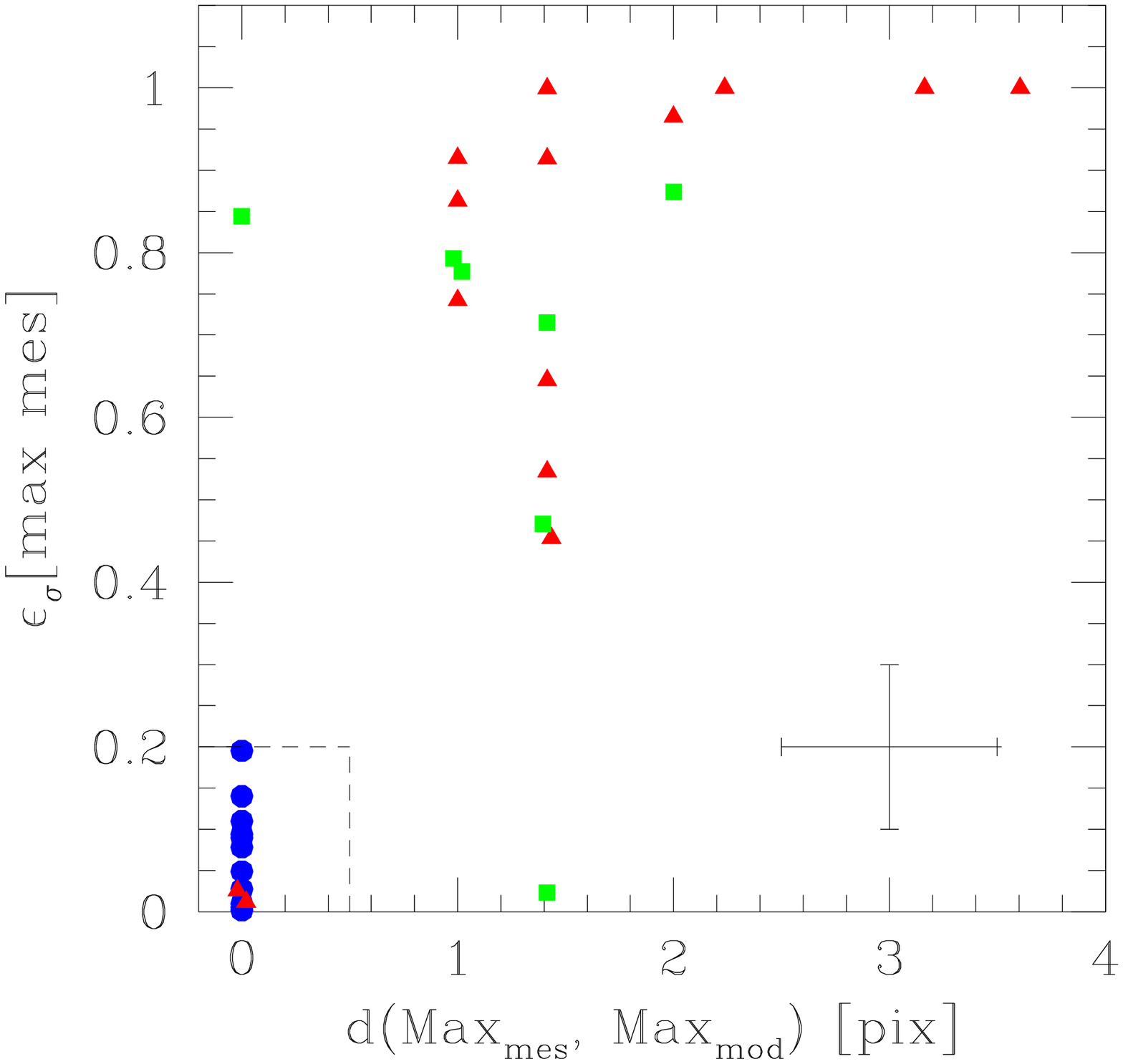}
\caption{ Plot $d(MAX_{obs},MAX_{mod})$ versus $\epsilon_\sigma[max
obs]$: blue dots, green squares and red triangles represent the
galaxies classified as rotation disk, perturbed rotation and complex
kinematics, respectively. All the galaxies classified as rotating disk
but one (CFRS031349) are concentrated in a small region of the plot
(see text for details). Median errorbars are indicated on the right of
the plot. The vertical errorbar varies from 5\% to 16\% ($median=
10$\%). The horizontal errorbar is estimated to half a GIRAFFE
pixel. }
\label{indices}
\end{figure}

\subsection{V$_{max}$ of distant galaxies}
\begin{figure}[h!]
\centering
\includegraphics[width=8cm]{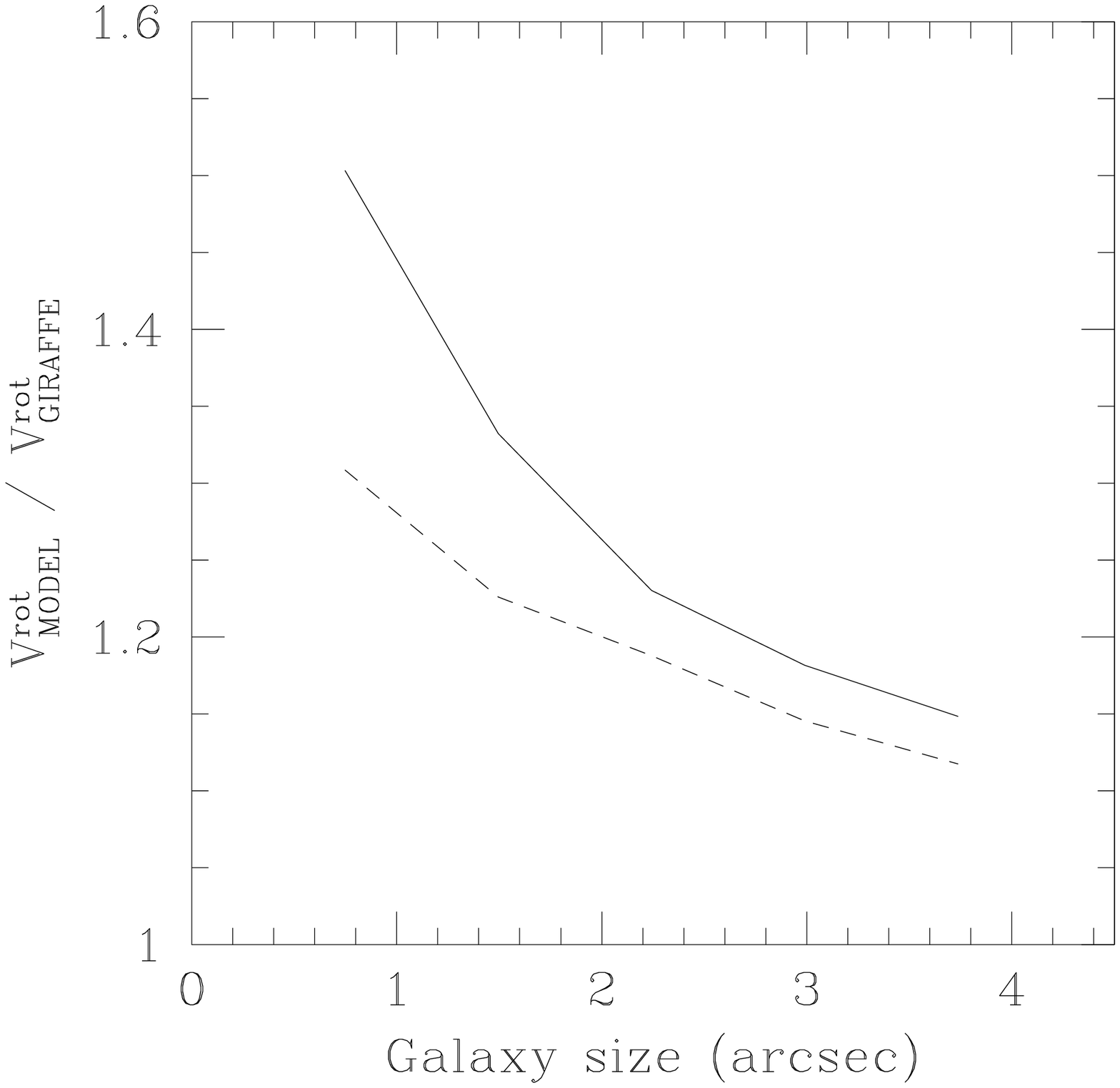}
\caption{Correcting factor for the maximal rotational velocity vs
galaxy size. Due to its poor spatial sampling, the GIRAFFE IFU leads
to underestimating $V_{rot}$ by a factor $1.20\pm 0.04$ for galaxies
sizes between 2 and 3 arcsec. Full line: tilted view ($\sim 53$
degrees). Dash line: top view ($\sim 25$ degrees) from Cox et al. (2004).}
\label{size}
\end{figure}

To determine V$_{max}$ for each galaxy we derive 2V$_{max}$ using the
direct measure of the maximal gradient of the velocity field from
FLAMES/GIRAFFE. Given the sizes of the galaxies and that of the IFU,
we are likely sampling an area that is large enough to recover
V$_{max}$ of the disk galaxy. Following Persic \& Salucci (1991),
V$_{max}$ is measured at r$_{half}$ $\ge$ 1.5R$_d$ (disk radius R$_d$, equals to 2.2
times the effective radius R$_e$). At z=0.55, GIRAFFE can measure V$_{max}$
for all the galaxies with r$_{half} <$ 8 kpc ($<$3 GIRAFFE spixals),
which is the case for all rotating disks of the sample (Table 2).
Using hydrodynamical simulations of a Sbc Milky Way-like galaxy by Cox
et al. (2004), we simulated GIRAFFE observations assuming median
atmospheric conditions at ESO VLT (0.81 arcsec seeing at 500 nm). We
scaled the template ($V_{rot}=160$ km/s and $i=53$ degrees) to fill
boxes of length ranging from 0.75 to 4 arcsec to mimic distant
galaxies, and then compared kinematics seen by GIRAFFE with the
original simulation (see Figure \ref{size}). We found that for spiral
galaxies with sizes ranging from 1.5 to 3 arcsec, GIRAFFE
underestimates the maximal rotational velocity by $\sim$ 20\%. For
more complex kinematics, the correction factor should be larger, given
the velocity gradient (see Figure 3 and discussion in Puech et al.,
2006a). In the following, we choose to assume a constant factor of
20\% independent dynamical class (spiral or perturbed/complex). 
This factor is roughly consistent with the one derived from the
simulations of Kapferer et al. (2005) mimicking long-slit spectroscopy
of distant rotating disks.

 One could derive individual correction factors from numerical
simulations using a very detailed modeling. However, this would
require a large number of parameters to be fitted, and given the small
number of spatial pixels in the GIRAFFE IFU, it would most probably
become difficult to solve the degeneracy between some of these
parameters. Thus, a possible residual error of 20\% (0.08 dex,
depending on the size of the galaxy) could affect the derivation of
the maximal velocity. Only 3D instruments with better spatial
resolution could give a better determination of V$_{max}$.

\section{The distant Tully-Fisher relation}
We have investigated the Tully-Fisher relationship of intermediate redshift galaxies using two absolute bands B and K, as well as an estimate of the stellar mass. Absolute magnitudes were derived following Hammer et al. (2001 and 2005), on the basis of interpolating the photometric measurements from B, V, I and K bands. Hammer et al. (2001) estimated that the global errors on the derived absolute magnitudes are comparable to the photometric error measurements, i.e. less than 0.1-0.2 magnitude (see Table 2). We then correct all absolute magnitude values by considering corrections related to the inclination-dependent intrinsic absorption due to dust. For this, we have followed Tully et al. (1998) in applying corrections that depend on the galaxy luminosity (see their equations 7 to 14). Following Tully et al. (2004), we applied an additional correction of 0.27 magnitude in the B band (0.04 mag in K band) which accounts for extinction correction applied to face-on galaxies (see also Tully \& Fouqu\'e 1985). 

The stellar masses of these galaxies have been estimated from K-band magnitudes and optical colors. We have chosen a conservative approach, which
assumes that  $M/L_K$  depends on the rest-frame B-V color following the  relation
derived by Bell et al. (2003; see Hammer et al. 2005 for more details). This approach empirically accounts for the contamination by young populations of super red giant stars, which can affect the K band luminosity, especially for young starbursts.

\subsection{K band TF relation}
Verheijen (2001) and Karachentsev et al. (2002) show a local TF relationship which is quasi-linear in K band with a slope ranging from -11.3 to -9.  Verheijen (2001) claims that the scatter of the relationship is much smaller than $\sigma$=0.17, a value provided by the intrinsic uncertainty of the distance of Ursae Major galaxies. As expected, the TF relationship in K band shows the smallest scatter when compared to that established at bluer wavelengths (Verheijen, 2001). 

\begin{figure}[h!]
\centering
\includegraphics[width=8cm]{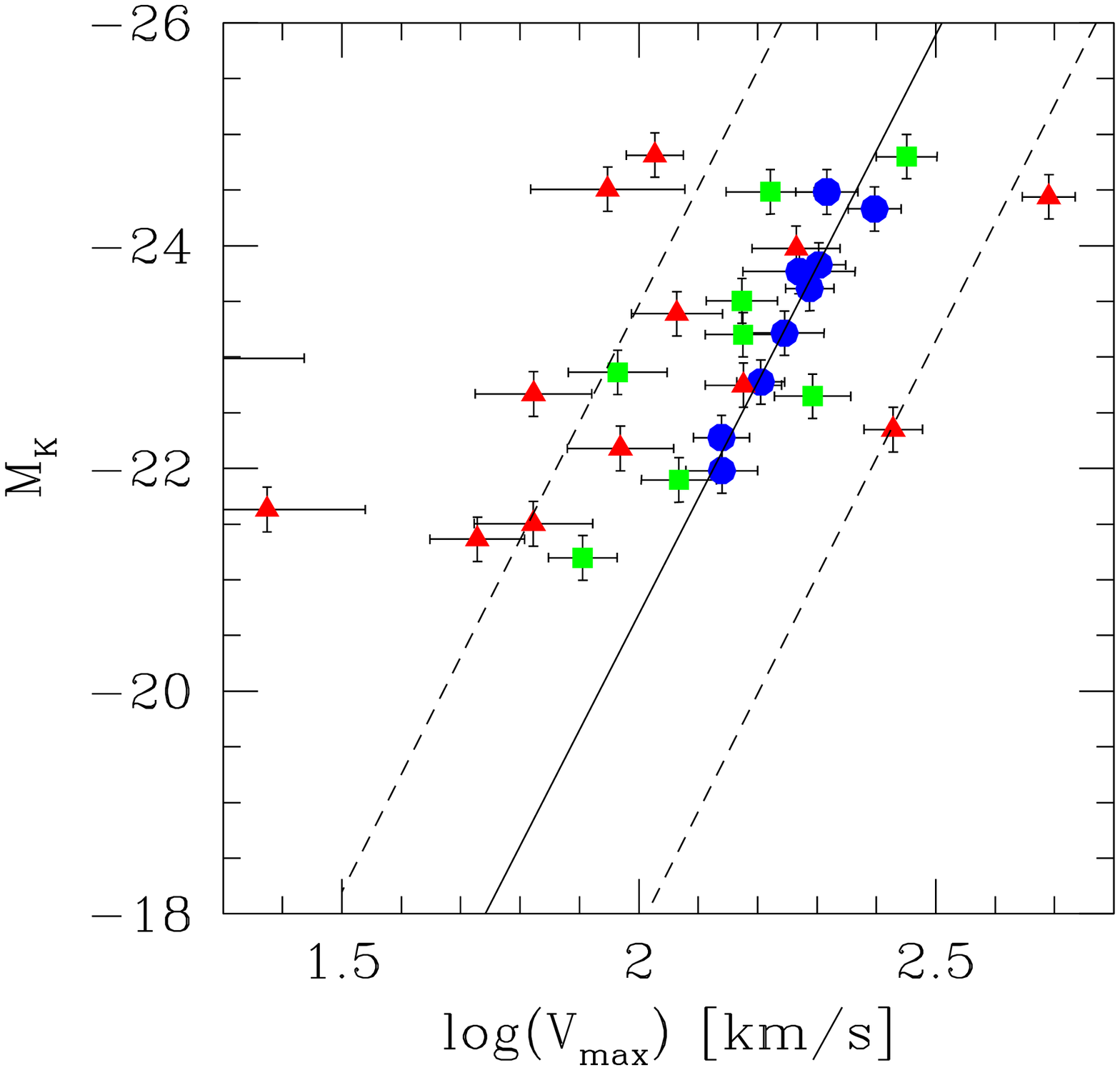}
\caption{TF relation in K band (Vega magnitudes). Red triangles, green squares and blue dots  represent complex kinematics, perturbed rotations and
rotating disks, respectively. Full and dotted lines represent the local TF and its 3 sigma scatter amplitude (following Conselice et al., 2005). This plot shows that all the scatter of
the TF relation is caused by interloper galaxies with kinematics 
classified either as complex or perturbed. Considering only rotating disks, the TF relationship at moderate redshift is similar (slope, zero point and scatter) to that in the local universe. For each galaxy, the error bars are shown.}
\label{FigTFK}
\end{figure}

Figure \ref{FigTFK} presents the TF relationship for 30 intermediate redshift galaxies for which we have been able to calculate their absolute K band magnitude (see Table 2). It does not differ considerably from that found by Conselice et al. (2005), except that it samples only the highest luminosity range. As in Conselice et al. (2005), it shows a very large scatter, which exceeds by several magnitudes what is found for local galaxies. Further examination of the relationship shows that almost all the scatter is related to those galaxies whose velocity fields have been classified either as complex or perturbed rotation. By considering only the 10 rotating disks, almost all the scatter is removed, and the TF relationship at moderate redshift is likely very similar (slope, zero point and scatter) to what has been found for local galaxies.

\begin{figure}[h!]
\centering
\includegraphics[width=8cm]{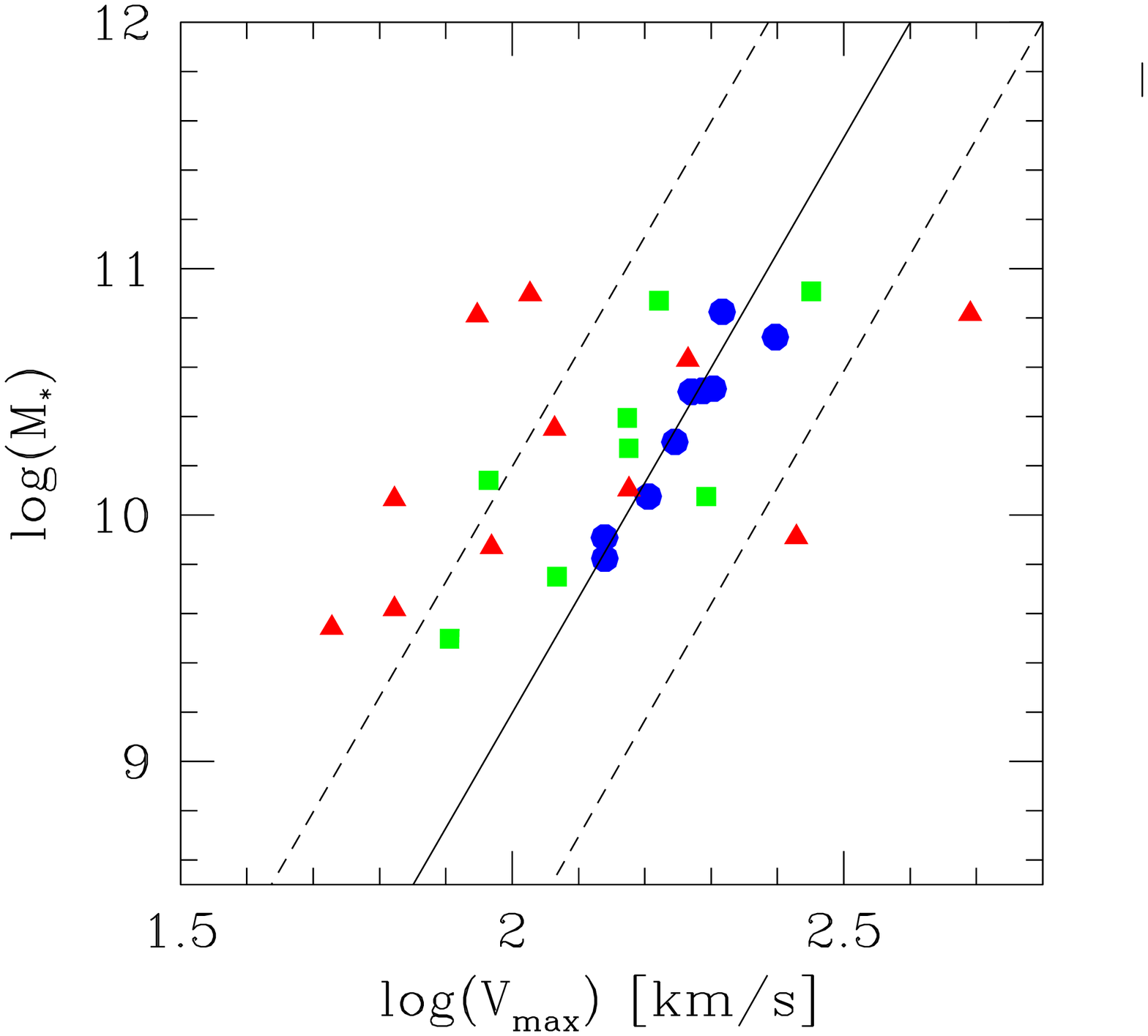}
\caption{Stellar mass versus V$_{max}$ relation of distant galaxies. Symbols arethe same as in Figure 6. The same effect is detected as in the K band. The relationship is again very similar to that found for local galaxies. Error bars on the stellar mass are typically 0.1-0.2 dex (see Hammer et al., 2005).}
\label{FigTFMass}
\end{figure}

A similar effect is found when considering the TF relationship with stellar masses (Figure \ref{FigTFMass}) instead of $M_K$. The relationship is again very similar to that found for local galaxies, with perhaps a slight trend that distant spirals at a given V$_{max}$ value show a lower stellar mass (on average by 0.1-0.2 dex) than local spirals. 

\subsection{B band TF relation}

\begin{figure}[h!]
\centering
\includegraphics[width=8cm]{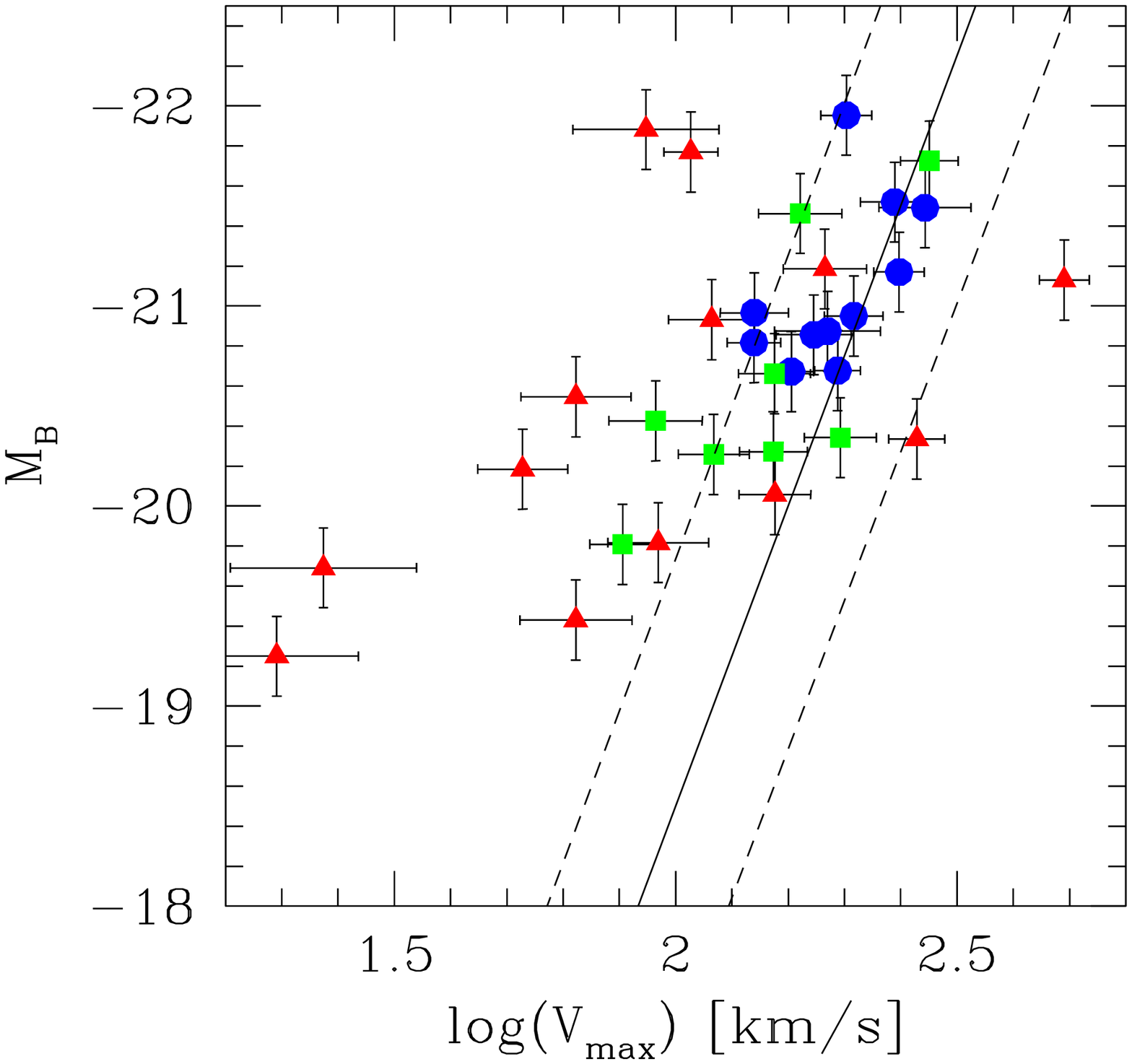}
\caption{TF relation in the B band (Vega magnitude, corrected for extinction). Symbols are the same as in Figure 6. Error bars for each galaxy are included. The scatter of local disks in this relation appear to be larger than in Figure 6 and 7, although  always smaller than the 3 sigma of the local relation , dotted lines) this could be related to an star formation event. }
\label{FigTFB}
\end{figure}

Figure \ref{FigTFB} presents the TF relationship based on the $M_B$ magnitude for 32 intermediate redshift galaxies. As in section 4.1, most of the dispersion is related to interlopers, i.e. to galaxies with velocity fields different to those of rotating disks. For the eleven rotating disks however we notice that the scatter is larger than what is found for local galaxies. We have verified whether this can be due to the uncertainty on the correction related to the inclination (which is much larger at B wavelength than that at K band), which could especially affect galaxies that have not been imaged by the HST. It seems that it is not the case because the three galaxies lying at 3 $\sigma$ above the local TF have HST images available (HDFS4020, 4180 and 5190). There are 3 to 5 spirals that show a higher B luminosity than that expected from the local TF, while none show the reverse effect (lower B luminosity than expected). It might be understood if we were directly viewing a highly active star formation event (such as a giant HII region) in some distant spirals. 

\section{Discussion}

Using the GIRAFFE multiple IFU modes available at VLT, we have been able to characterise the velocity fields of 32 intermediate redshift galaxies (z$\sim$ 0.55) among 35 galaxies that have been selected for their apparent magnitudes ($I_{AB}$ $<$ 22.5) and their emission line properties ($W_0$([OII]) $>$ 15 \AA$\;$). This sample, although small,  is representative of the general properties of emission line galaxies selected in the Canada France Redshift Survey (see Hammer et al., 1997). Indeed, the GIRAFFE wavelength range is so small (approximately 800\AA$\;$ for the LR04 and LR05) that the number of available objects in the field is equal to or lower than the available number of IFUs, so our selection is unlikely to provide strong biases.

Our classification scheme is very simple, assuming the velocity field of a rotating disk to be a standard (class "rotating disk", RD). A discrepancy  in the $\sigma$ map (absence of a peak in the centre of an apparently rotating velocity field) leads us to classify the kinematics as "perturbed rotation" (class PR), since it might illustrate a possible minor merger event that does not affect the disk stability, but has an expected signature in the $\sigma$ map. We have compared our observed $\sigma$ map to a simple perfect rotating disk model. All systems showing two or more discrepancies in either the velocity field maps or in the $\sigma$ maps ( "complex kinematics" class, CK) are discrepant. In our next paper, further work will be dedicated to improve these classifications, including by the differenciation of the various complex kinematics
This will be useful, especially when a larger set of data become available.
 
 We could have misidentified some observed velocity fields, and it could affect our conclusions, especially for the 6 velocity fields for which the kinematical properties of the [OII] doublet have been characterised by only 6 to 8 spatial GIRAFFE pixels (i.e. less or equal to 2 spatial resolution elements, see Table 2). Among them, 3 are classified as "perturbed rotation" and 3 as "complex kinematics".  Further investigation of their individual spectra in the IFU casts some doubt about the classification of 2 galaxies (CFRS031016 and HDFS4090) for which the kinematics has been classified as complex (CK). Indeed all show a well resolved [OII] doublet, a property shared by most rotating disks in our sample, but generally not by objects with complex kinematics. The complete sample of 35 galaxies with $I_{AB}$ $<$ 22.5 and $W_0$([OII]) $>$ 15 \AA includes  3 objects that are so compact that the emission line area
  is too small to conclude on their nature. Thus, it is difficult to conclude on the kinematics of 5 (2+3) objects in the complete sample, 4 of them being compact galaxies. In this sample, the fraction of rotating disks ranges from 1/3 (11/35) to 1/2 (18/35) depending on the final classification of the above 5 galaxies with uncertain velocity fields. Because these are related to compact galaxies, which are dominated by complex kinematics (see Table 3 and Puech et al., 2006a), it is reasonable that the rotating disk fraction is closer to 1/3 of the sample. 

May be our sample is contaminated by low mass interlopers, since Figure 7 shows that rotating disks have slightly higher stellar masses than the average. By limiting our sample to  $M_{star}$ $>$ $10^{10}$ $M_{\odot}$, 7 rotating disks are found within 21 galaxies (33\%), so this cannot affect our conclusions. A similar but stronger effect can be found in Figure 8, which shows that rotating disks are brighter in B than other galaxies. This might be interpreted as a "downsizing" effect, i.e. the less massive or luminous objects show more evolution. It might also be understood as a subtle selection effect, because it is probable that relaxed systems, such as rotating disks, are less efficient in producing stars than systems with complex or perturbed kinematics. To be observed with GIRAFFE, galaxies must have enough [OII] flux, and we suspect that some rotating disks with low masses might have been missed due  to our selection criterion.

\section{Conclusion}

We find that 3D spectroscopy is the only way to probe the dynamics of distant galaxies and the evolution of the Tully-Fisher relationship. FLAMES/GIRAFFE, with its system of 15 deployable IFUs, is uniquely able to perform these measurements. All the trends described below have to be confirmed using a larger data set of distant galaxy velocity fields, scheduled in the framework of a large program at VLT entitled IMAGES (Intermediate MAss Galaxy Evolution Sequence).

\begin{enumerate}
\item {\bf Almost half of the emission line galaxies at moderate redshifts have complex kinematics, revealing mergers, merger remnants and inflow/outflows.}
This shows that random large scale motion  dominates their velocity fields. Obvious mergers seen by imagery (e.g. CFRS031309) share this property as do most of the compact galaxies, and this support the Hammer et al. (2005) conjecture that merging is, at moderate redshifts, an important process in activating star formation. The 7 objects (22\%) with perturbed rotation ($\sigma$ peak off the center) could be the result of minor mergers or interactions (e.g. CFRS031349). Those would not destroy the disk rotation but are expected to increase the dispersion near the impact area or to significantly distort the sigma peak. Inflow/outflows motion may be revealed by a large offset between the optical and the dynamical axis. This could be the case for 4 galaxies (12\% of the sample)  which have compact morphologies and generally have low stellar masses (see Puech et al., 2006a). In such cases, it is possible that these motions dominate the rotational motions. 
In most cases, complex kinematics is suggestive of mergers or merger remnants for which the circular motion in the disk has been destroyed, leading to large amounts of gas motion as revealed by GIRAFFE. This will be investigated in a forthcoming paper by comparing our observations to realistic numerical simulations such as those performed by Cox et al. (2004).

\item {\bf At z$\sim$ 0.6, almost 40\% of all galaxies with $M_B$ $<$ -19.5 are not rotating disks, i.e. they are not at dynamical equilibrium.} The fraction of intermediate galaxies that has not reached a dynamical equilibrium is very large:  66\% of of non rotating disks in our sample corresponding to 36\% of all z$\sim$ 0.6 galaxies with $M_B$ $<$ -19.5. This is based on Hammer et al. (1997) who found that 60\% of z$\sim$ 0.6 galaxies have $W_0$([OII]) $>$ 15 \AA$\;$. Interestingly, our finding that at moderate redshifts, 36\% of galaxies are  not rotating disks or ellipticals supported by dispersion corresponds to the fraction of peculiar galaxies (irregular/compact/merger, see Zheng et al., 2004, 2005 and references therein), while those objects are almost absent in samples of nearby galaxies. We also notice that most galaxies (but not all) with unambiguous spiral morphology show rotational motion, a property shared by some irregulars (e.g. CFRS039003). Table 3 also reveals a trend between kinematic stability and the half light radius, since less than 1/4 of compact galaxies are rotating disks. 

\item {\bf We find that the previously reported evolution of the Tully-Fisher relationship is due to instrumental causes,} i.e. to the incapacity of slit spectroscopy to distinguish rotating disks from other perturbed or complex rotational fields. Our work resolves the origin of the large scatter found by all previous studies based on slit spectroscopy. Selecting a sample at random within the field emission line galaxies would lead to only 34\% being rotating disks, that could be used to establish a proper Tully-Fisher relation. Even by selecting extended objects (r$_{half}$ $>$ 4.7 kpc), the number of objects contaminated by random motion is still large (47\%).   It is unclear whether this prevents the use of slits in deriving the TF of distant galaxies. 

\item {\bf The Tully-Fisher relationship (stellar mass-velocity) has apparently not evolved, either in slope, zero point or scatter.}  Rejecting the galaxies with non-relaxed dynamics leads, at z$\sim$ 0.6, to a Tully-Fisher relationship that shows no evidence of a redshift change when comparing velocity to $M_K$ or stellar mass. It somewhat strengthens our confidence in our estimates of stellar mass, since the rotating disks actually display the whole range of star formation activity from starbursts to LIRGs. An absence of evolution of the Tully-Fisher relationship with stellar mass will provide in the near future an important tool to distinguish predictions from galaxy formation scenarios. Its reported evolution in the blue shows that $\sim$1/3 of the spirals undergo significant brightening of their blue luminosities, as expected if they are hosting active star formation. Interestingly a similar effect has been found in a sample of galaxies in compact groups (see Mendes de Oliveira et al., 2003).

\end{enumerate}

\begin{acknowledgements}
We thank all the GIRAFFE teams at the Paris Observatory, Geneve
Observatory and ESO for this unique instrument, without which, none of
these results would have been obtained. We thank A. Bosma for his
enlightening comment on CFRS030508. We are especially indebted to
T.J. Cox who provided us with hydrodynamical simulations of an Sbc
galaxy. C. Ravikumar helped us with the English. HF and MP thank ESO
Paranal staff for their support and useful advices. The referee,
Richard Bower is warmly acknowledged for the useful and constructive
comment.
\end{acknowledgements}

\end{document}